\documentstyle[12pt,epsf]{article}

\newcommand{\beq}{\begin{equation}}
\newcommand{\eeq}{\end{equation}}
\newcommand{\bea}{\begin{eqnarray}}
\newcommand{\eea}{\end{eqnarray}}
\newcommand{\rar}{\rightarrow}
\newcommand{\lra}{\longrightarrow}
\newcommand{\lan}{\langle}
\newcommand{\ran}{\rangle}

\begin{document}
\font\fortssbx=cmssbx10 scaled \magstep2
\hbox to \hsize{
\hskip.5in \raise.1in\hbox{\fortssbx University of Wisconsin - Madison}
\hfill$\vcenter{\hbox{\bf MADPH-96-938}
                      \hbox{June 1996}}$ }
\vskip 2cm
\begin{center}
\Large
{\bf Sum rules, Regge trajectories, and relativistic quark models} \\
\vskip 0.5cm
\large
Sini\v{s}a Veseli and M. G. Olsson\\
\vskip 0.1cm
{\small \em Department of Physics, University of Wisconsin, Madison,
	\rm WI 53706}
\vspace*{+0.2cm}
\end{center}
\thispagestyle{empty}
\vskip 0.7cm

\begin{abstract}
We present an analysis which uses Regge structure
and the Bjorken and Voloshin
HQET sum rules to restrict the choice of
parameters of a relativistic quark model.
\end{abstract}

\newpage
\section{Introduction}
It is a well known fact that almost any relativistic quark model
which involves a reasonable quark-antiquark potential
will adequately reproduce the spin-averaged spectrum
of the heavy-heavy and heavy-light
states. Almost all successful potentials are based on some
variant of a one gluon exchange (Coulomb) part plus a
confining part expected from QCD. However, the specific choice of
parameters of the model is usually based only on their ability
to reproduce data. That and the fact that the parameters
of the model are correlated as far as meson spectrum
is concerned,\footnote{By this we mean that changing one
parameter in the model
inevitably leads to changes in other parameters. For example,
it is a well known fact that it is much
easier to determine the difference between $b$ and $c$
quark mass from the meson spectrum, than it is to
determine  either mass.}
is the main reason why one can find nearly as
many different sets of parameters for the same model,
as there are papers using that particular model.

In this letter we advocate an analysis which uses the
linear Regge structure
of a given relativistic quark model
in the light-light limit \cite{regge,goebel}, together with
sum rules of the heavy quark effective theory
in the heavy-light limit \cite{bjsr,vosr}, in order
to constrain the parameters of a given potential.
For definiteness, we use the simplest and the most
widely used generalization of the nonrelativistic
Schr$\ddot{\rm o}$dinger equation
\cite{durand1}-\cite{jacobs},
the so called spinless Salpeter or the square root  equation.
It is clear though that
the same reasoning can be applied to any relativistic quark model
which exhibits linear Regge behavior in the light-light limit.

The aim of this paper is not to say whether a particular
model is right or wrong in terms of the type of confinement or the
particular  potential it uses. Instead, our goal is
to show how one can construct a model which is consistent with
experiment in terms of its Regge structure in the light-light limit,
and at  the same time self-consistent with respect to the sum rules in
the  heavy-light limit.

In Section \ref{model} we give a brief description of
the model we are using as an example of our analysis.
Section \ref{rt} discusses the  implications of the linear
Regge structure of the model in the light-light limit.
Sum rule constraints on the model are presented in
Section \ref{sumr}, while our conclusions are summarized in
Section \ref{conc}.

\section{Relativistic quark model}
\label{model}

As already mentioned, for definiteness we use the
relativistic quark model with Hamiltonian given by
\cite{durand1}-\cite{jacobs}
\beq
H = \sqrt{m_{1}^{2} + p^{2}} + \sqrt{m_{2}^{2} + p^{2}} + V(r)
\ ,
\label{ham}
\eeq
where $p^{2} = p_{r}^{2} + l(l+1)/r^{2}$,
$V(r)= V_{conf}(r) + V_{oge}(r)$, and
\bea
V_{oge}(r) &=& -\frac{4}{3}\frac{\alpha_{s}}{r}\ ,
\label{voge}\\
V_{conf}(r) &=& b r + c\ .
\label{vconf}
\eea
The necessity of adding a (negative) constant to the usual linear confining
potential was shown in \cite{gromes}, from the non-relativistic
limit of the Bethe-Salpeter equation.
There it was emphasized that $c$ is a parameter which is as
fundamental and indispensable as the quark masses, slope of the
linear potential $b$, and the strong coupling constant
$\alpha_{s}$.\footnote{
Under additional assumptions an even stronger result
$c\simeq -2\sqrt{b}\exp{(-\gamma_{E} + 1/2})$ was obtained in \cite{gromes}.
We shall, however, consider
$c$ as an independent parameter.}

One can find many papers\footnote{References
\cite{durand1}-\cite{jacobs} and \cite{lucha}-\cite{hwang}
are just a few of them.
Probably the most ambitious and the most
sophisticated version of the model is due to Godfrey and Isgur
\cite{godfrey}.} in the literature
which use the above model
with or without relativistic corrections (e.g. the spin-orbit
and color hyperfine interaction) which are completely specified
in terms of quark masses and parameters of $V_{conf}(r) $
and $V_{oge}(r)$,  with the fixed effective or with some sort of running
coupling constant. For the sake of simplicity, we use an
effective short range coupling constant, since there is little this
analysis can say about $V_{oge}(r)$.

As one can see from (\ref{ham}), (\ref{voge}) and (\ref{vconf}),
in order to completely specify the model (which should be able
to reproduce spin-averaged heavy-heavy and heavy-light meson states)
we need seven parameters (we assume that the $u$ and $d$
quarks have the same mass): four constituent quark masses ($m_{u,d},
m_{s}, m_{c}$ and $m_{b}$), $\alpha_{s}$, and two parameters
specifying the confining part of the potential, $b$ and $c$.
Again, there is little one can say about quark masses or $\alpha_{s}$,
since these are expected to run as one goes from the
light-light to  the heavy-light and heavy-heavy systems. However, we can
try to constrain  the confining part of the potential so that the
model is at least self-consistent in the light-light
and the heavy-light limits.

\section{Regge trajectories}
\label{rt}

Let us first consider the effective string tension $b$.
It is a well known experimental fact that all light-quark hadrons
lie upon linear Regge trajectories with a universal
slope \cite{barger}.
While it may have been already pointed out in the past
\cite{regge,goebel} we would like to reemphasize here the relation
between linear trajectories, linear confinement, and relativistic dynamics.
It seems inescapable that massless quarks bound by a linear
confinement potential generate a family of parallel linear Regge
trajectories, whose slopes depend on the Lorentz nature and
other properties of the interaction.

The leading Regge slope follows
from the correspondence (classical) limit. The lowest energy state
of (\ref{ham})
for a given large angular momentum
results for circular orbits at large $r$ and $p$. The minimal energy
condition $\left.\frac{\partial H}{\partial r}\right|_{L}\equiv 0$
implies that ($p_{r}\rar 0, \ p\rar L/r$)
\beq
Lp(\frac{1}{E_{1}} + \frac{1}{E_{2}}) = b r^{2}\ ,
\label{l1}
\eeq
where $E_{i} = \sqrt{p^{2} + m^{2}_{i}}$.
For a light-light meson $m_{1,2}\rar 0$, $E_{1,2}\rar p$, and hence
(\ref{l1}) gives $2L=  b r^{2}$. This, together with  $H \rar 2L/r + b r$,
implies that Regge slope  is given by \cite{goebel}
\beq
\alpha'_{LL} = \frac{L}{H^{2}} =  \frac{1}{8 b}\ .
\eeq
Combining this with the observed slope of the leading $\rho$ trajectory
\cite{barger,montanet},
\beq
\alpha'^{exp}_{LL} = 0.88\ GeV^{-2}\ ,
\label{sl1}
\eeq
we see that in order for the model to be consistent with experiment in the
light-light
limit, we have to require
\beq
b = 0.142\ GeV^{2}\ .
\label{b}
\eeq
This requirement is often overlooked in the literature,\footnote{
A most recent example is Ref. \cite{hwang}.
There, (\ref{ham}) was used with several different potentials, and
they  were
all inconsistent with the Regge structure of the model in the
light-light limit.}
and $b$ is
usually fixed to be about $0.18\ GeV^{2}$ (see Table \ref{tabconf} for
a few examples), which is
a value suitable for models  with the Nambu string slope of
$\alpha'_{LL} = 1/(2 \pi b)$ (e.g. flux tube models \cite{rft}).

A similar result for the effective string tension $b$
can also be obtained
from the Regge behavior of the model in the heavy-light limit,
since
it is expected that the Regge slope for the energy of
the light degrees of freedom
is twice the slope for the light-light case \cite{goebel}, i.e.
$\alpha'_{HL} = 2 \alpha'_{LL}$.
Indeed, in our case for a heavy-light meson we have
$E_{2}\rar m_{2}\rar \infty$
and $E_{1}\rar p$ ($m_{1}\rar 0$), and hence (\ref{l1}) gives
$L = br^{2}$. Together with $H\rar m_{2} + L/r + b r$, this
yields
\beq
\alpha'_{HL} = \frac{L}{(H-m_{2})^{2}} = \frac{1}{4b}\ .
\label{sl2}
\eeq
We should also note that the results
(\ref{sl1}) and (\ref{sl2}) depend on the nature of the
confining potential used with (\ref{ham}), and that the above arguments
cannot be used with models which do not
exhibit linear Regge behavior (e.g. non-relativistic quark model).

\section{HQET sum rules}
\label{sumr}

Let us now consider the heavy-light limit of (\ref{ham}),
\beq
H\stackrel{m_{2}\rar\infty}{\lra} m_{2} + \sqrt{m_{1}^{2}+p^{2}}+V(r)\ .
\eeq
It is clear that in this case the constant $c$ can be reabsorbed
into the heavy quark mass by $m_{2}\rar m_{2} - c$, and that the
heavy-light spin-averaged meson spectrum by itself does not contain
enough information    to  determine $c$. However,
additional constraints on the value of $c$ can be obtained from the
Bjorken \cite{bjsr} and Voloshin \cite{vosr} sum rules.

In the heavy-light limit
the angular momentum  of the light degrees of freedom (LDF)
decouples from the spin of the heavy quark, and both
are separately conserved by the strong interaction. Therefore, the
total angular momentum $j$ of the LDF  is a good quantum number. For
each $j$ there are two
degenerate heavy-light states ($J=j\pm \frac{1}{2}$), which
can be labeled  as $J^{P}_{j}$. Let us denote  the
$L = 0$ doublet ($0^{-}_{\frac{1}{2}}$ and $1^{-}_{\frac{1}{2}}$ states) by
($C$, $C^{*}$), and by
($E$, $E^{*}$) and  ($F$, $F^{*}$) the two
$L=1$ doublets
 (($0^{+}_{\frac{1}{2}}$, $1^{+}_{\frac{1}{2}}$) and
($1^{+}_{\frac{3}{2}}$, $2^{+}_{\frac{3}{2}}$), respectively).
Correspondingly, the unknown form factors for the semileptonic
decays of the $B$ (or $B_{s}$) meson are $\xi_{C}$ (for $C\rar (C,C^{*})$),
$\xi_{E}$ (for $C\rar (E,E^{*}$)), and $\xi_{F}$
(for $C\rar (F,F^{*})$ transitions).
They are defined within the covariant trace formalism \cite{cft}, and
are functions of the four-velocity transfer $\omega = v\cdot v'$, where
$v$ and $v'$ denote four-velocities of the initial and final heavy-light
state.
In terms of these form factors
the Bjorken sum rule \cite{bjsr} is given by
\beq
-\xi'_{C}(1) = \frac{1}{4} +
\frac{1}{4}\sum_{i}\left| \xi_{E}^{(i)}(1)\right|^{2}
+\frac{2}{3}\sum_{j}\left| \xi_{F}^{(j)}(1)\right|^{2}\ .
\label{bjsr}
\eeq
The sums here are understood in a generalized
sense as sums over discrete states and integrals over continuum states.
Similarly, the Voloshin sum rule \cite{vosr}
can be written as
\beq
 \frac{1}{2}
=
\frac{1}{4}\sum_{i}(\frac{E_{E}^{(i)}}{E_{C}}-1)
\left| \xi_{E}^{(i)}(1)\right|^{2}
+\frac{2}{3}\sum_{j}(\frac{E_{F}^{(j)}}{E_{C}}-1)
\left| \xi_{F}^{(j)}(1)\right|^{2}\equiv \Delta\ .
\label{vosr}
\eeq
In the above expressions $E_{E}^{(i)}$ and $E_{F}^{(j)}$
denote energies of the LDF in the $i$-th excited state
with  quantum numbers of the $(E,E^{*})$ doublet, and
$j$-th excited state with quantum numbers of the $(F,F^{*})$
doublet, respectively. $E_{C}$
is the LDF energy in the lowest ($C,C^{*}$) doublet (corresponding
to ($D,D^{*}$) or ($D_{s},D_{s}^{*}$) mesons). The energy
of the LDF in any
heavy-light state  is defined as the state mass
minus the heavy quark mass.

Since in our model we cannot say anything about
continuum contributions to these two sum rules, and since all
contributions in the sums of (\ref{bjsr}) and (\ref{vosr})
are positive definite, one can argue
that resonant contributions to the right-hand sides of
(\ref{bjsr}) and (\ref{vosr}) should be smaller
than direct calculation of $-\xi'_{C}(1)$ and $1/2$,
respectively, if a model is to be consistent with
these two sum rules. This will in fact be the key argument of
the sum rule part of the model analysis.

In the valence quark approximation one can find
precise definitions of the form factors which appear
in  expressions (\ref{bjsr}) and (\ref{vosr}) (consistent with
the covariant trace formalism \cite{cft}), in terms of the wave functions
and energies of the LDF \cite{modelling}. We reproduce below
only the expressions for $\xi_{C}'(1)$,
$\xi_{E}(1)$ and $\xi_{F}(1)$, which  are necessary for the
sum rule analysis of a model:
\bea
\xi'_{C}(1) &=& -\frac{1}{2} - \frac{1}{3}
E_{C}^{2}\lan r^{2}\ran_{00}\ ,\label{xicsl}
\\
\xi_{E}(1)& =& \frac{1}{3}(E_{C}+E_{E})\lan r\ran_{10}\ ,
\label{xie}\\
\xi_{F}(1)& =& \frac{1}{2\sqrt{3}}
(E_{C}+E_{F})\lan r\ran_{10}\ .
\label{xif}
\eea

Since the model we are considering is spinless, the $E$ and $F$ doublets
are degenerate and the
expectation values in (\ref{xicsl})-(\ref{xif}) are  defined as
\beq
\lan F(r) \ran_{L'L}^{\alpha'\alpha} = \int r^{2} dr R^{*}_{\alpha'L'}(r)
R_{\alpha L}(r)F(r)\ ,
\label{jave}
\eeq
where $R_{\alpha L}(r)$ are radial wave functions of a heavy-light
system.
Also,
(\ref{xie}) and (\ref{xif}) in the spinless case imply
that
\beq
\xi_{E}(1) = \frac{2}{\sqrt{3}}\xi_{F}(1)\ .
\label{xief}
\eeq
Using (\ref{xie}) and (\ref{xif}), one can  simplify the
expressions for both sum rules. The Bjorken sum rule (\ref{bjsr}) becomes
\beq
-\xi'_{C}(1) \geq \frac{1}{4} +
\frac{1}{12}\sum_{i}\left[(E_{C}+E_{E,F}^{(i)})\lan r\ran_{10}^{(i)}\right]^{2}
\ ,
\label{bjsr2}
\eeq
while the Voloshin sum rule (\ref{vosr}) is now given by
\beq
 \frac{1}{2}
\geq
\frac{1}{12}\sum_{i}(\frac{E_{E,F}^{(i)}}{E_{C}}-1)
\left[(E_{C}+E_{E,F}^{(i)})\lan r\ran_{10}^{(i)}\right]^{2}
\equiv \Delta\ .
\label{vosr2}
\eeq
In these two equations the sums are only over resonances, since
we are neglecting continuum  contributions.

To understand the essence of the sum rule constraint we first
consider a slightly oversimplified example.
We assume that non-resonant contributions
are small, and then the equality sign holds in
(\ref{bjsr2}) and (\ref{vosr2}). We also assume  that the
sums in these two equations are saturated by the lowest
$P$-wave doublets, then (\ref{bjsr2}) and (\ref{vosr2})
imply
\bea
-\xi'_{C}&\simeq& \frac{1}{4} + \frac{1}{12}(2E_{C}+\Delta E)^{2}
\lan r\ran_{10}^{2} \ ,
\label{bjsr3}\\
\frac{1}{2}&\simeq & \frac{1}{12}\frac{\Delta E}{E_{C}}
(2 E_{C}+ \Delta E)^{2}\lan r\ran_{10}^{2}\ .
\label{vosr3}
\eea
In the above equations we have written
$E_{E,F}^{(1)} = E_{C} + \Delta E$, where $\Delta E$ is just
the difference between the spin-averaged masses of the
$S$-wave and the lowest $P$-wave doublet.\footnote{If one assumes that the
spin-averaged mass
of the lowest ($E,E^{*}$)
doublet in the $D$ systems  (corresponding to $D_{0}$ and $D_{1}$ mesons),
is about
2350 $MeV$, then this difference is about
440 $MeV$.} From (\ref{vosr3}) one obtains
\beq
(2 E_{C}+\Delta E)^{2}\lan r\ran_{10}^{2} \simeq
6\frac{E_{C}}{\Delta E}\ .
\eeq
Using the above expression in (\ref{bjsr3}), together
with the bound $-\xi'_{C}(1)\geq 1/2$ from (\ref{xicsl}),
we find
\beq
E_{C}\geq \frac{1}{2}\Delta E\ .
\eeq
This implies that energy of the LDF in the lowest $S$-wave
mesons cannot be smaller than one half of the difference
between the spin-averaged masses of the lowest $P$-wave and the $S$-wave.
In particular, this would also set a lower bound on
the constant $c$, which is usually assumed to be negative,
or an upper bound on the constituent heavy quark masses, since
the energy of the LDF is defined as the state mass minus the
heavy quark mass.

We now go back to the model analysis of (\ref{bjsr2})
and (\ref{vosr2}). As already mentioned, for the
description of heavy-light mesons we need seven
parameters. We have already determined the effective string tension
$b$ from the
consistency of the model with the experimental
Regge slopes of light-light mesons. In order to determine
all other parameters except $c$, we can use the observed
heavy-light spectrum.
Any change in $c$ effectively just changes
the heavy constituent  quark masses by the same amount, and an
equivalent description of the spectrum is obtained. It is evident that
the heavy-light wave functions are not affected by this change. However,
changing $c$ does affect the energies of the LDF in the heavy-light mesons.
This in turn affects the form factors predicted by the model.
The idea is that by examining the model predictions
for the right-hand sides of (\ref{bjsr2}) and (\ref{vosr2}),
one can determine physically acceptable values of $c$, and
other parameters of the model. As explained earlier, the
main requirement here is self-consistency of the
model in the sense that its predictions for the
right-hand sides of (\ref{bjsr2}) and (\ref{vosr2})
are smaller than its direct
calculation of $-\xi'_{C}(1)$,
and $1/2$, respectively.
However, in order to account for all
possible  uncertainties in our calculations (e.g. effects of
spin-averaging of $P$-wave states, assumption of the unknown
$P$-wave masses, etc.), we relax the sum rule constraints
by $5\%$. This means that the sum rule calculation of $\Delta$ should
yield result smaller than 0.525 (instead of 0.5), and that the sum rule
calculation of $-\xi'_{C}(1)$ should yield a result which is at most
$5\%$ larger than the result obtained from the direct calculation.
In this way a more conservative bounds on $c$ will be obtained.

In order to illustrate the above ideas, we consider two sets of parameters:
\begin{itemize}
\item Set 1: we fix $c = 0$ and  $m_{u,d} = 350\ MeV$, and a fit
to the spin-averaged heavy-light meson spectrum then yields
$m_{s} = 542 \ MeV$, $m_{c} = 1366\ MeV$,
$m_{b} = 4703\ MeV$, and $\alpha_{s} = 0.390$.
\item Set 2: we fix $c = 0$ and  $m_{u,d} = 300\ MeV$,
and from the fit to the heavy-light data we then obtain
$m_{s} = 503 \ MeV$, $m_{c} = 1390\ MeV$,
$m_{b} = 4726\ MeV$, and $\alpha_{s} = 0.390$.
\end{itemize}
For both of these two sets we used
$b  = 0.142\ GeV^{2}$ from (\ref{b}), and both of them
 yield an excellent
description of the known heavy-light (spin-averaged)
meson masses, with errors less than $5\ MeV$.\footnote{We have
assumed that
$D_{0}$ ($0^{+}_{\frac{1}{2}}$ state) and $D_{1} $
($1^{+}_{\frac{1}{2}}$ state) have spin-averaged mass of
$2350\ MeV$, which together with the known $P$ waves
$D_{1}(2425)$ and $D_{2}^{*}(2459)$ leads to  the
spin-averaged mass of $2414\ MeV$ for the lowest $P$-wave
in  the $D$ systems.
Heavy quark symmetry arguments then imply
a spin-averaged mass of $2523\ MeV$ for the corresponding
lowest $P$-wave in the $D_{s}$ systems.}
Using the parameters which
reproduce spin-averaged data, and also an effective
string tension consistent with experiment, gives
us confidence that the unknown spin-averaged meson masses
for radial excitations
are reproduced reasonably well.

For both of these two sets, and for $c$ ranging
from $0$ to $-600\ MeV$, we have evaluated $-\xi'_{C}$
directly using (\ref{xicsl}).
Using 5 lowest
$P$-waves we also evaluated the right-hand sides
of (\ref{bjsr2}) and (\ref{vosr2}). The results of our calculations
(for $B \rar D,D^{*}$ semileptonic decays)
are shown in Figures  \ref{bjsr_sseq} (for the Bjorken sum rule)
and \ref{vosr_sseq} (for the Voloshin sum rule).
As one can see in  Figure \ref{bjsr_sseq}, both sets of parameters
can satisfy the Bjorken sum rule in the sense that the direct calculation
of $-\xi'_{C}(1)$ yields a larger result than the sum rule approach.
Imposing the weaker requirement, i.e. that the sum rule result is at most
$5\%$ larger than the direct result, for set 1 we find
$c\leq -180\ MeV$, while for set 2 we get $c\leq -185\ MeV$. We also note
that for these values of $c$,  $-\xi'_{C}(1)\simeq 0.71$ (for set 1) and
$-\xi'_{C}(1)\simeq 0.70$ (for set 2).
For the Voloshin sum rule (Figure \ref{vosr_sseq})
we find that parameter set 1 can satisfy the stronger constraint
($\Delta \leq 0.5$) for some values of $c$, which is not
the case for  parameter set 2.
However, imposing the 5\% relaxed bound ($\Delta \leq 0.525$), we
find that  model is self-consistent with
\beq
200\ MeV \leq -c \leq 470\ MeV\ ,
\label{rangec}
\eeq
for parameter set 1, and with
\beq
265\ MeV \leq -c \leq 410\ MeV\ ,
\label{rangec2}
\eeq
for parameter set 2.
Therefore, in the two cases considered, we found
the Voloshin sum rule to be more restrictive than
the Bjorken sum rule.
It is  also  interesting to observe that the minimum of the
function $\Delta(c)$ always occurs  close to the point at which
Bjorken sum rule approach yields $-\xi'_{C}(1)=1/2$.
That can be shown analytically from (\ref{bjsr3}) and (\ref{vosr3}),
and can be seen from Figures \ref{bjsr_sseq} and
\ref{vosr_sseq}.\footnote{This is not necessarily the case
with models based on the Dirac (or Salpeter) equation.}
For the constituent heavy quark masses values of $c$
given in (\ref{rangec}) imply
 $1566\ MeV \leq m_{c} \leq 1836\ MeV$, and
$4903\ MeV \leq m_{b} \leq 5173\ MeV$ (set 1), while
the ones given in (\ref{rangec2}) yield
 $1655\ MeV \leq m_{c} \leq 1800\ MeV$, and
$4991\ MeV \leq m_{b} \leq 5136\ MeV$ (set 2).
Also, the range of energy of the LDF in the $S$-wave
$(C, C^{*}$) heavy-light
meson corresponding to (\ref{rangec}) is
$140\ MeV \leq E_{C} \leq 410\ MeV$ (set 1), while
the one corresponding to (\ref{rangec2}) is
$175\ MeV \leq E_{C} \leq 320\ MeV$ (set 2).
These values for $E_{C}$ are to be compared with  the ones
obtained in \cite{onepar}, where it was found
$266\ MeV \leq E_{C} \leq 346\ MeV$.
This result followed from the analysis of the
recent data on semileptonic $B$ decays \cite{barish} using the
$1S$ lattice QCD heavy-light wave function \cite{duncan}.
We have also repeated the same calculation for the
$B_{s} \rar D_{s},D_{s}^{*}$ semileptonic decays
with the same
sets of parameters  (the difference
is basically only the light quark mass), and
found that these decays are much less restrictive than
corresponding $B\rar D,D^{*}$ decays.

The most serious concern which one might have about
our sum rule analysis of a model, is
the issue of degeneracy of the two $P$-wave doublets, which
is due to the spinless nature of the particular model we
considered in this paper.\footnote{For models based on the
Dirac or Salpeter equation this issue vanishes, since
these models distinguish between $(E,E^{*})$ and $(F,F^{*})$
doublets. Because of that, sum rule analysis of these models
should be more reliable than the one appropriate for the spinless
models.}
While it is certainly true that
the uncertainties introduced in this way should partly
cancel out (due to spin-averaging, contributions of
some states to sum rules will be overestimated, and for other states
they will be underestimated), we have still tried to account
for possible theoretical errors by relaxing the sum rule constraints
by 5\%. Also note that including more radially excited states
would yield more strict bounds on the
acceptable values of $c$,
than are the ones we quote.
Given all that, we believe that
(\ref{rangec}) and (\ref{rangec2}) represent reasonably
conservative estimates for the two different parameter  sets of
the model considered
in this paper. This conclusion is also supported by the
comparison of the values for $E_{C}$ obtained here
with the ones obtained in \cite{onepar}.

\section{Conclusion}
\label{conc}

In the construction of potential
models of mesons a confining interaction of the form
$V_{conf}(r) = b r + c$ plays an important role.
The  additive constant $c$ in a number of cases just renormalizes
quark masses, but has very little effect on meson wave function.
On the other hand, form factors for the semileptonic
decays of $B$ (or $B_{s}$) meson in the heavy quark limit
depend sensitively on this constant through the energies
of the light degrees of freedom. The result is a wide
variety of form factor predictions from a class of very similar models.

Using the simplest and the most
widely used generalization of the nonrelativistic
Schr$\ddot{\rm o}$dinger equation \cite{durand1}-\cite{jacobs},
we have presented an analysis,
based on the linear Regge structure of the
model in the light-light limit,
and its predictions for the two HQET sum rules
in the heavy-light limit,
which can constrain parameters of the confining
part of the potential. Even though we have used a particular
relativistic quark model as an example, it is clear
that a similar analysis could be repeated for any model
with linear Regge behavior.

The aim of this
analysis is not to show whether a particular
model is correct in terms of a given equation,
a type of confinement, or a specific potential it uses.
We have instead outlined a method to restrict the
confining parameters of the model in order to
make it consistent with experiment in the light-light limit
(Regge structure),  and also self-consistent with respect
to HQET sum rules in  the heavy-light limit.
Once these parameters are fixed, other parameters
(e.g. constituent quark masses and short range coupling
constant) can be determined from the requirement
that model should reproduce various
experimental data (e.g. meson masses).

\begin{center}
ACKNOWLEDGMENTS
\end{center}
This work was supported in part by the U.S. Department of Energy
under Contract No. DE-FG02-95ER40896 and in part by the University
of Wisconsin Research Committee with funds granted by the Wisconsin Alumni
Research Foundation.

\clearpage
\newpage

\begin{table}
\begin{center}
TABLE
\vskip 2mm
\end{center}
\caption{Parameters of the confining part of the potential
used in several papers which employed the relativistic
generalization of the Schr$\ddot{\rm o}$dinger equation discussed
in the text, with the same type of the confining potential
(\protect\ref{vconf}).
The one gluon exchange potentials used in these papers are not
necessarily the same as (\protect\ref{voge}).
Reference \protect\cite{fulcher} used two different values for $c$
for description of $D$ and $B$ mesons.
To obtain the universal  Regge slope model
requires $b=0.142\ GeV^{2}$, while the
range of the $c$ values for which  model
is consistent with HQET sum rules depends on other assumptions
and parameters of the model. }
\label{tabconf}

\begin{center}
\begin{tabular}{|lccc|}
\hline
\hline
Author(s) & Reference   &    $b\ [GeV^{2}]$   &   $-c\ [MeV]$ \\
\hline
Durand and Durand (1984) & \protect\cite{durand2} & 0.180  & 0 \\
Godfrey and Isgur (1985) & \protect\cite{godfrey} & 0.180  & 253 \\
Jacobs et. al. (1986)    &\protect\cite{jacobs} & 0.192  & 0 \\
Lucha et. al.  (1992)    & \protect\cite{lucha} & 0.211  & 850 \\
Fulcher et. al.  (1993)  & \protect\cite{fulcher} & 0.191  & 246 (214) \\
Fulcher          (1994)  & \protect\cite{fulcher2} & 0.219  & 175 \\
Hwang and Kim    (1996)  & \protect\cite{hwang} & 0.183  & 0 \\
\hline
\hline
\end{tabular}

\end{center}
\end{table}

\clearpage

\newpage

\begin{figure}
\begin{center}
FIGURES
\vskip 2mm
\end{center}
\caption{Comparison of the direct calculation
of $-\xi'_{C}(1)$ (full lines) with the Bjorken sum rule result
obtained with 5 lowest $P$-wave states
(dashed lines). 1 and 2
denote two different sets of parameters, as explained in the text.
The dotted line is the bound  $-\xi'_{C}(1)\geq \frac{1}{2}$
coming from (\protect\ref{xicsl}).  The results shown are for
$B\rar D, D^{*}$ semileptonic decays.}
\label{bjsr_sseq}
\end{figure}

\begin{figure}
\caption{Voloshin sum rule
calculation of $\Delta$ with 5 lowest
$P$-wave states (full lines), for the two different
sets of parameters 1 and 2.
The expected upper bound of $0.5$
and the $5\%$ relaxed bound of $0.525$ are shown with dotted and
dashed line, respectively.
The results shown are for
$B \rar D, D^{*}$ semileptonic decays.}
\label{vosr_sseq}
\end{figure}

\begin{figure}
\vskip 6cm
\end{figure}

\begin{figure}[p]
\epsfxsize = 5.4in
\centerline{\vbox{\epsfbox{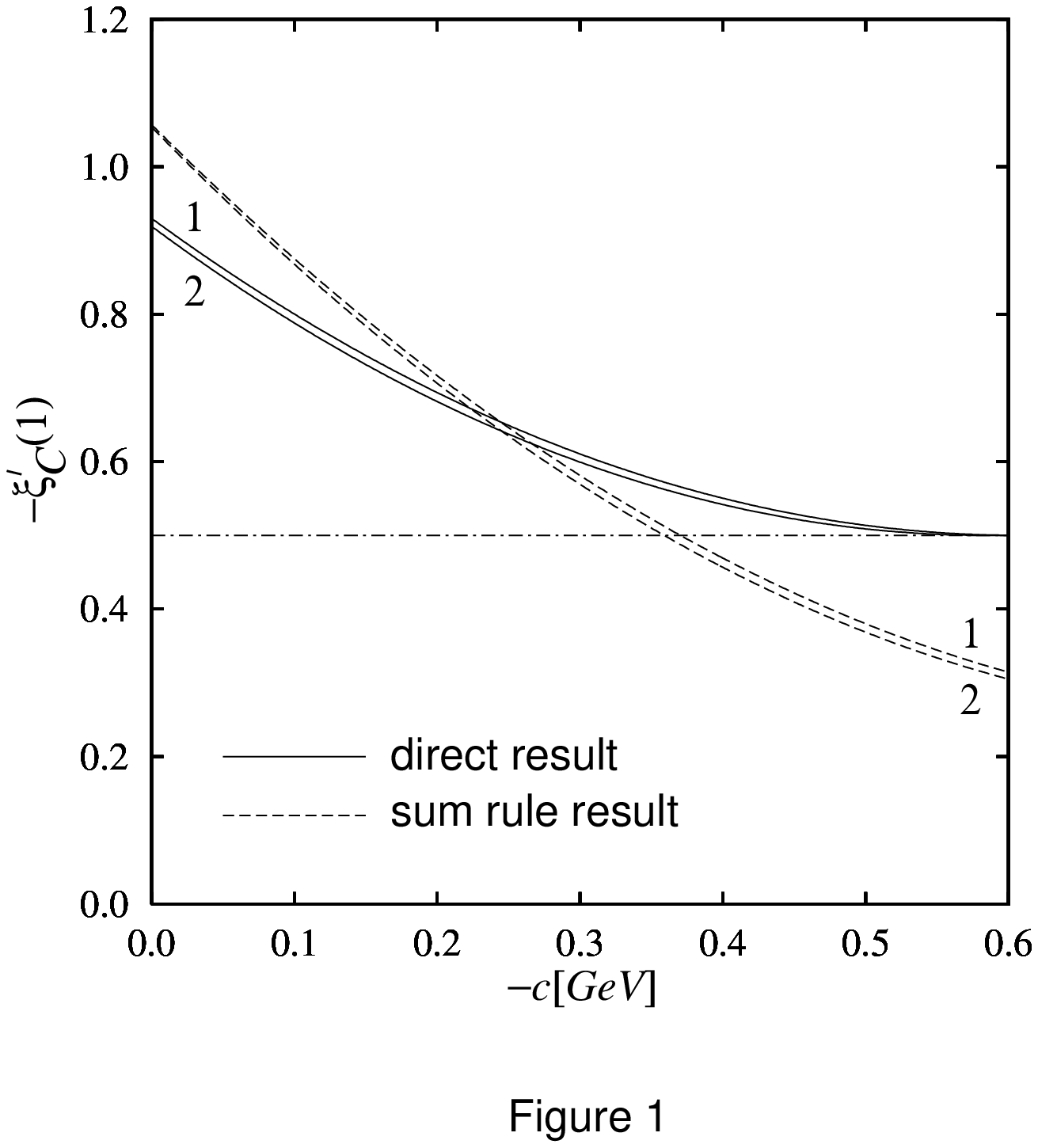}}}
\end{figure}

\begin{figure}[p]
\epsfxsize = 5.4in
\centerline{\vbox{\epsfbox{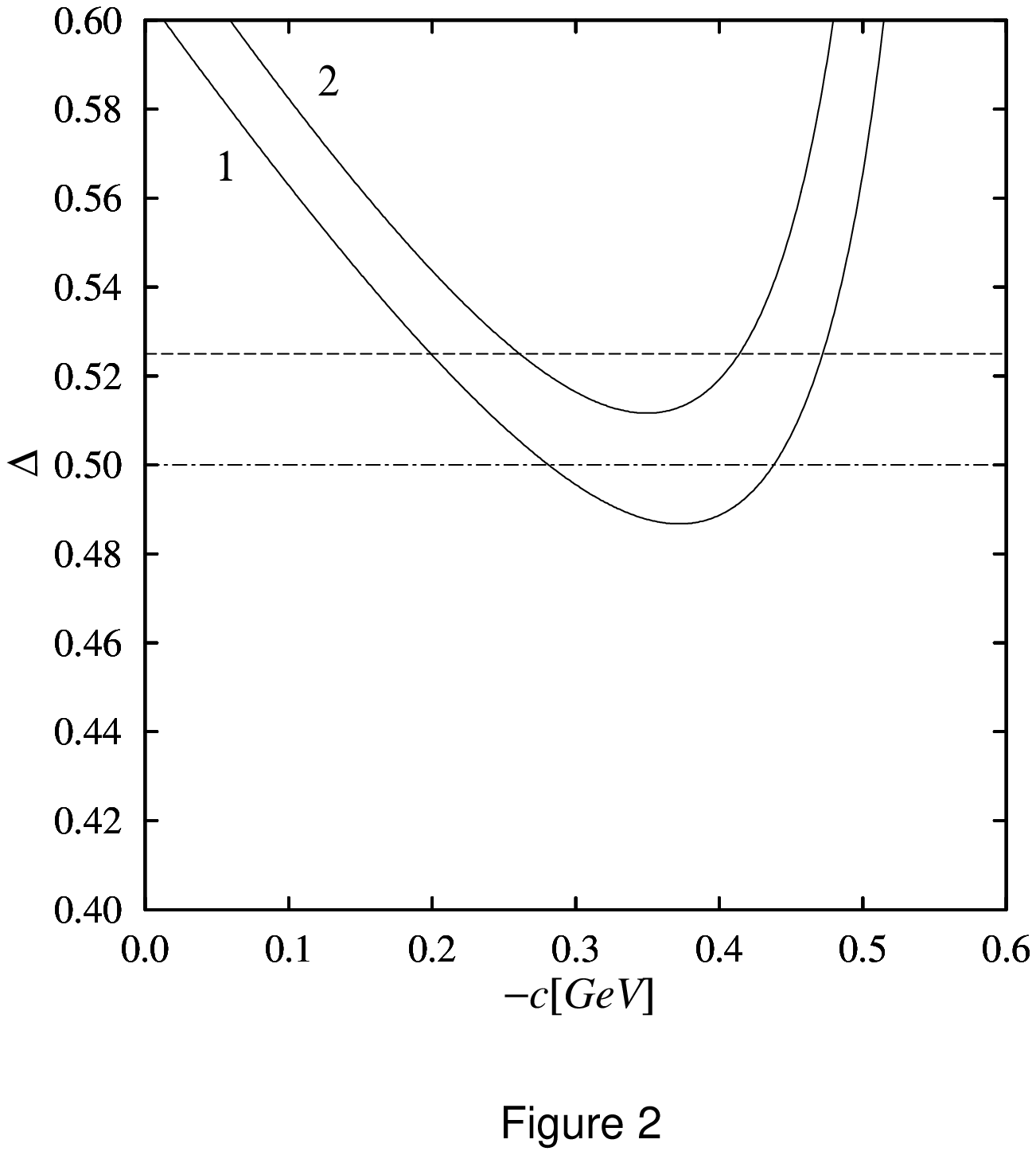}}}
\end{figure}

\end{document}